\documentclass[twocolumn,pra,aps,showpacs]{revtex4}
\usepackage[dvips]{graphicx}
\usepackage{graphicx}% Include figure files
\usepackage{dcolumn}% Align table columns on decimal point
\usepackage{bm}% bold math
\begin{document}

\title{Goos-H\"{a}nchen shifts in frustrated total internal reflection studied with wave packet propagation}

\author{Xi Chen$^{1,2}$\footnote{Email address: xchen@shu.edu.cn}}

\author{Chun-Fang Li$^{1,3}$}

\author{Rong-Rong Wei$^{1}$}

\author{Yan Zhang$^{4}$}

\affiliation{$^{1}$ Department of Physics, Shanghai University,
200444 Shanghai, People's Republic of China}

\affiliation{$^2$ Departamento de Qu\'{\i}mica-F\'{\i}sica, UPV-EHU,
Apdo 644, 48080 Bilbao, Spain}

\affiliation{$^{3}$ State Key Laboratory of Transient Optics and
Photonics, Xi'an Institute of Optics and Precision Mechanics of CAS,
710119 Xi'an, People's Republic of China}

\affiliation{$^{4}$ Department of Electronics and Information
Engineering, Shanghai University, 200072 Shanghai, People's Republic
of China}

\date{\today}

\begin{abstract}

We have investigated that the Goos-H\"{a}nchen (GH) shifts in
frustrated total internal reflection (FTIR) studied with wave packet
propagation. In the first-order approximation of the transmission
coefficient, the GH shift is exactly the expression given by
stationary phase method, thus saturates an asymptotic constant in
two different ways depending on the angle of incidence. Taking
account into the second-order approximation, the GH shift always
depends on the width of the air gap due to the modification of the
beam width. It is further shown that the GH shift with second-order
correction increases with decreasing the beam width at the small
incidence angles, while for the large incidence angles it reveals a
strong decrease with decreasing the beam width. These phenomena
offers the better understanding of the tunneling delay time in FTIR.



\end{abstract}

\maketitle
%----------------------------------------------------------------

It is well known that a light beam totally reflected from an
interface between two dielectric media undergoes lateral shift from
the position predicted by geometrical optics \cite{Goos}. This
phenomenon was referred to as Goos-H\"{a}nchen (GH) effect and was
theoretically explained by Artmann's stationary phase method
\cite{Artmann} and Renard's energy flux method \cite{Renard}.
Because of the potentials applications in integrated optics
\cite{Lotsch}, optical waveguide switch \cite{Sakata}, and optical
sensors \cite{Yin-Hesselink,Yu}, the GH shifts including other three
non-specular effects such as angular deflection, focal shift and
waist-width modification have been extensively investigated in
partial reflection \cite{Hsue-T,Riesz-S,Tamir,Li,Nimtz,Li-ZCZ},
attenuated total reflection \cite{Yin,Pillon}, and frustrated total
internal reflection (FTIR) \cite{Cowan,Agudin,Ghatak,Haibel,Broe}.

From a somewhat different perspective, the optical tunneling
phenomenon in FTIR have attracted much attention in the last two
decades \cite{Steinberg-C,Stahlhofen,Balcou,Haibel-Nimtz,Hooper},
because of the analogy between FITR and quantum tunneling.
Theoretical \cite{Steinberg-C,Stahlhofen} and experimental
\cite{Balcou,Haibel-Nimtz} investigations have demonstrated that the
GH shifts in FTIR play an important role in the superluminal
tunneling time and the well-known ``Hartman effect" \cite{Hartman},
which describes that the group delay for quantum particles tunneling
though a potential barrier saturates to a constant for an opaque
barrier. Recently, Martinez and Polatdemir \cite{Martinez} have
studied the effect of width of the beam on the GH shift (which is
proportional to tunneling time) to offer the complementary insights
into the origin of ``Hartman effect" in FTIR. In addition, Haibel
\textit{et al.} \cite{Haibel} once carried out a comprehensive study
of the GH shift in FTIR as a function of the polarization, beam
width, and incidence angle in the microwave experiment, which
challenges its theoretical descriptions. However, the current
expressions of the GH shifts given by stationary phase method and
energy flux method are independent of the beam width.

The main purpose of this Brief Report is to investigate that the GH
shifts in FTIR by wave packet propagation. It is shown that the GH
shift in the first-order approximation of the transmission
coefficient is exactly the expression of the GH shift given by
stationary phase method. The GH shift in this case approaches the
saturation value in two different ways depending on the incidence
angle. Taking account into the second-order approximation, the GH
shift always depends  on the width of air gap. It is further shown
that the GH shift with the second-order correction become dependent
strongly on the beam width. These phenomena offers the better
understanding of the tunneling delay time in FTIR.

\begin{figure}[]
\scalebox{0.55}[0.55]{
  \includegraphics{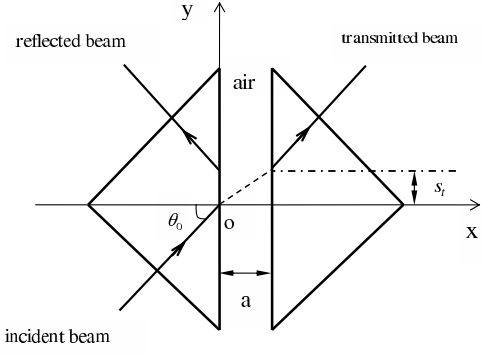}}
\caption{Schematic diagram of the GH shift in FTIR, where the width
of the air gap is $a$, $s_t$ represents the GH shift of the
transmitted beam.} \label{fig.1}
\end{figure}

For simplicity, we consider TE polarized beam incident into the
double-prism structure with the angular frequency $\omega$ and
incidence angle $\theta_0$, as shown in Fig. \ref{fig.1}, where $a$
is the width of the air gap. Denote by $\varepsilon$, $\mu$ and $n$,
respectively, the permittivity, permeability and refractive index of
the prism. For a well-collimated beam, the electric field of the
incident beam can be expressed by
\begin{equation}
\Psi_{in}(x)=\frac{1}{\sqrt{2\pi}}\int_{-\infty}^{\infty}A(k_y)\exp[i(k_xx+k_yy)]dk_y,
\end{equation}
where $k_x= n k\cos\theta$, $k_y= n k \sin\theta$, $k=\omega /c$,
$n=\sqrt{\epsilon\mu}$, $c$ is the speed of light in vacuum,
$\theta$ is the incident angle of the plane wave component under
consideration, and time dependence $\exp (-i \omega t)$ is implied
and suppressed. For a Gaussian-shaped incident beam whose peak is
assumed
to be located at $x=0$,
\begin{equation}
\label{incident beam}
\Psi_{i}(x=0,y)=\exp{\left(-\frac{y^2}{2w^2_y}\right)}\exp{(i
k_{y0}y)},
\end{equation}
its angular spectral distribution is also a Gaussian function,
$A(k_y)=w_y \exp[-(w_y^2/2)(k_y-k_{y0})^2],$ around its central
$k_{y0}=k \sin \theta_0$, $w_y=w_0/\cos\theta_0$, $w_0$ is the width
of the beam at waist. According to Maxwell equations and boundary
conditions, the field of the transmitted beam is found to be
\begin{equation}
\label{transmitted beam}
\Psi_{t}(x)=\frac{1}{\sqrt{2\pi}}\int_{-\infty}^{\infty}T
A(k_y)\exp\{i[k_x(x-a)+k_yy]\}dk_y,
\end{equation}
with the transmitted coefficient $T=\exp(i\phi)/f$ is given by
\begin{equation}
f\exp(i\phi)=\cosh\kappa{a}+ \frac{i}{2}
\left(\frac{k_x}{\kappa}-\frac{\kappa}{k_x}\right) \sinh \kappa{a},
\end{equation}
where $\kappa=(k^2_y-k^{2})^{1/2}$.

Firstly, we look at the GH shift in the first-order approximation of
the transmission coefficient. Expand the exponent in Taylor series
at $k_{y0}$, and retain up to the first-order term, then we will
obtain
\begin{equation}
\label{transmission series} T(k_y) = \exp[\ln T(k_y)] \approx T_0
\exp{\left[\frac{1}{T_0}\frac{dT}{dk_{y0}}(k_y-k_{y0})\right]},
\end{equation}
where $T_0=T(k_{y0})$, and $d/dk_{y0}$ denotes the derivative with
respect to $k_y$ evaluated at $k_y=k_{y0}$. Introduce two real
parameters $L_t'$ and $L_t''$ defined as,
\begin{equation}
 L = L_t'+iL_t'' = \frac{i}{T_0}\frac{dT}{dk_{y0}},
\end{equation}
then, in terms of the phase and magnitude of $T$, we will have
$$L_t'=-\frac{d\phi}{dk_{y0}},$$
and
$$L_t''=\frac{d}{dk_{y0}}\ln|T(k_y)|.$$
Substituting Eq. (\ref{transmission series}) into Eq.
(\ref{transmitted beam}) and employing the paraxial approximation
condition,
\begin{equation}
\label{approximation condition} k_x \approx
k_{x0}-(k_y-k_{y0})\tan{\theta_0}- \frac{(k_y-k_{y0})^2}{2k \cos^2
\theta_0},
\end{equation}
we obtain the transmitted beam at $x=a$,
\begin{equation}
\label{field of transmitted beam} \Psi_t(a,y) \approx T_0
\exp{\left[-\frac{(y-L_t')^2}{2 w^2_y}\right]} \exp{\left[i
(k_{y0}+\frac{L_t''}{w^2_y})y\right]},
\end{equation}
It is clear that the lateral shift $L_t'= - d\phi/dk_{y0}$ is the
same as the one obtained by the stationary phase method
\cite{Artmann}, and is given by
\begin{eqnarray}
\label{GH shift} s^p_t &=& \frac{s_c}{2
f_0^2}\left[\left(\frac{\kappa_0}{k_{x0}}-\frac{k_{x0}}{\kappa_0}\right)\right.
\\\nonumber &+&
\left.\left(\frac{k_{x0}}{\kappa_0}+\frac{\kappa_0}{k_{x0}}\right)\left(1+\frac{\kappa^2_0}{k^2_{x0}}\right)\frac{\sinh
2 \kappa_0 a}{2 \kappa_0 a} \right],
\end{eqnarray}
where $s_c =a k_{y0}/\kappa_0$. When the width of the air gap is
large enough, that is, $a \gg 1/\kappa$, the GH shift will tends to
a constant,
\begin{equation}
s^p_{asymp} \equiv \lim_{\kappa_0 a \rightarrow
\infty}{s^p_t}=\frac{2k_{y0}}{\kappa_0{k_{x0}}}.
\end{equation}
With increasing the air gap the GH shift reaches a asymptotic
constant, which is in agreement with the experimental results
\cite{Haibel}, and is also closely related to the counterintuitive
``Hartman effect" of the tunneling delay time in the limit of an
opaque barrier \cite{Balcou,Haibel-Nimtz}.

More interestingly, what we emphasized here is that the GH shift
approaches the saturation value in two different ways depending on
the angle of incidence. The GH shift (\ref{GH shift}) can be
expressed by the following form:
\begin{eqnarray}
 s^p_t &=& \frac{g_0}{1+g^2_0}\left[\left(1+\frac{\kappa_0^2}{k_{x0}^2}\right)\left(1+\frac{k_{x0}^2}{\kappa_0^2}
 \right)\frac{k_{y0}}{k_{x0}^2- \kappa_0^2}\right.
 \nonumber \\&&- \left.\frac{2ak_{y0}}{\kappa_0{\sinh{2\kappa_0
 a}}}\right],
\end{eqnarray}
where
$g_0=(k_{x0}^2-\kappa_0^2)\tanh{{\kappa_0}a}/{2k_{x0}\kappa_0}$.
Keeping the next to leading term for large $a$ shows the approach to
asymptotic value by
\begin{equation}
 s^p_t \simeq s^p_{asymp}+ 8 a
 \left(\frac{k_{y0}}{k_{x0}}\right)\left[\frac{k^2_{x0}(\kappa^2_0-k^2_{x0})}{(\kappa^2_0+k^2_{x0})^2}\right]e^{- 2 \kappa_0
 a}.
\end{equation}
It is clearly evident from the above expression that for
$\kappa^2_0-k^2_{x0}<0$ the GH shift increases monotonically to
reach the saturation value, while for $\kappa^2_0-k^2_{x0}>0$ it
reaches the saturation value from above, that is, there is a hump
before it attains saturation. Therefore, when the necessary
condition for incident angle,
\begin{equation}
\theta_0 > \theta_p \equiv \sin^{-1}\sqrt\frac{1+n^2}{2n^2},
\end{equation}
is satisfied,
\begin{figure}[]
\scalebox{0.30}[0.30]{
  \includegraphics{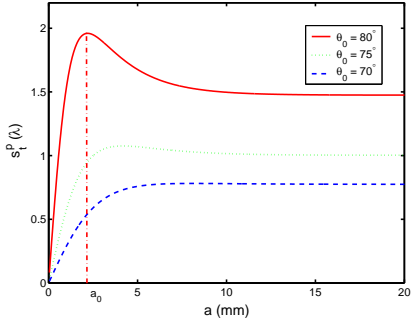}}
\caption{(Color online) Dependence of the GH shifts (in unit of
$\lambda$) on the air gap width $a$ in FTIR, where $\lambda=32.8
\mbox{mm}$ and $n=1.605$, the solid, dotted and dashed curves
correspond to the incidence angles $80^{\circ}$, $75^{\circ}$, and
$70^{\circ}$, respectively.} \label{fig.2}
\end{figure}
the GH shift can approach the saturation limit with negative slope
and can be larger than the saturation value for the intermediate
values of the air gap. The position of the maximum of the hump
($a_0$) is given by
\begin{equation}
a_{0}=\frac{1}{2\kappa_0}\frac{\kappa_0^4+3k_{x0}^2\kappa_0^2}{(k_{x0}^2+\kappa_0^2)^2}.
\end{equation}
Fig. \ref{fig.2} shows that for the large $a$ the GH shift $s^p_t$
is independent of the width $a$ of the air gap hence it saturates a
asymptotic constant, where $\lambda=32.8 \mbox{mm}$ and $n=1.605$
(corresponding to critical angle $\theta_c = 38.5^{\circ}$ for total
reflection and $\theta_p = 56.4^{\circ}$) \cite{Haibel}.
Furthermore, the GH shift approaches the asymptotic limit from above
for $\theta_0 > \theta_p$ and from below for $\theta_0 < \theta_p$.
These phenomenon is not due to the interference time
\cite{Martinez}, and does result from the interference between the
incident and reflected beams. Of course, it can also been seen from
the relationship between the GH shift and group delay discussed in
Ref. \cite{Winful-Zhang} that the delay time in FTIR also saturates
to a constant from above for $\theta_0 > \theta_p$ \cite{Caculation}
in the same way as that in the quantum tunneling for $E<V_0/2$
\cite{Solli,Paul}, since the self-interference delay time that comes
from the overlap of incident and reflected waves in front of barrier
is of great importance \cite{Winful}.

Next, what as follows we will show the influence of the beam waist
width on the GH shift in FTIR. To this end, we consider the exponent
of the transmission coefficient is approximated to the second-order
term,
\begin{eqnarray}
\label{transmission series-2} T(k_y) &\approx& T_0 \exp
\left[\frac{1}{T_0}\frac{dT}{dk_{y0}}(k_y-k_{y0}) \right.
\\\nonumber &+& \left.
\frac{1}{2}\frac{d}{dk_{y0}}\left(\frac{1}{T}\frac{d
T}{dk_y}\right)(k_y-k_{y0})^2\right].
\end{eqnarray}
Introducing two new real parameters $F_t^{'}$ and $F_t^{''}$ defined
as
\begin{figure}[]
\scalebox{0.22}[0.25]{\includegraphics{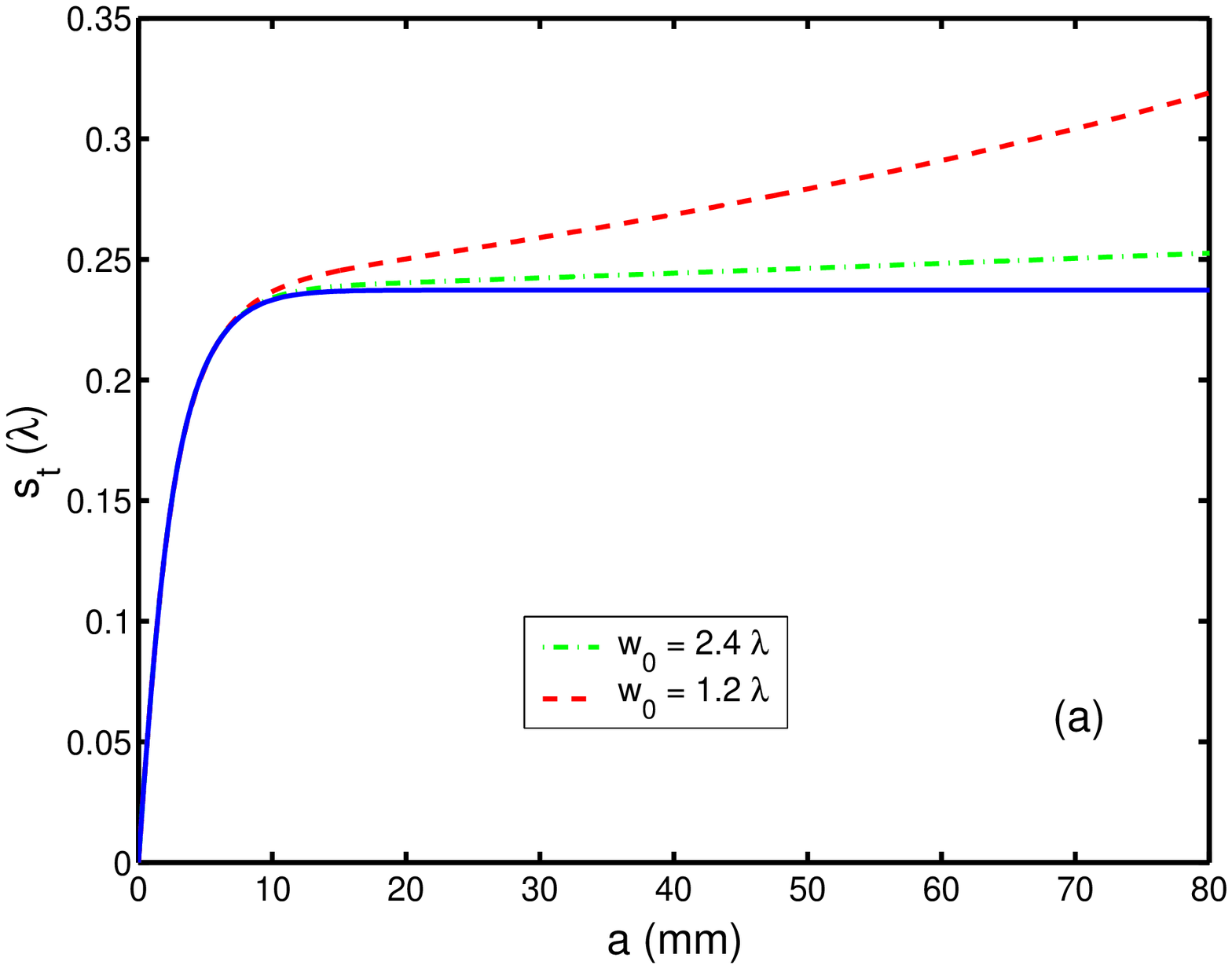}}
\scalebox{0.22}[0.25]{\includegraphics{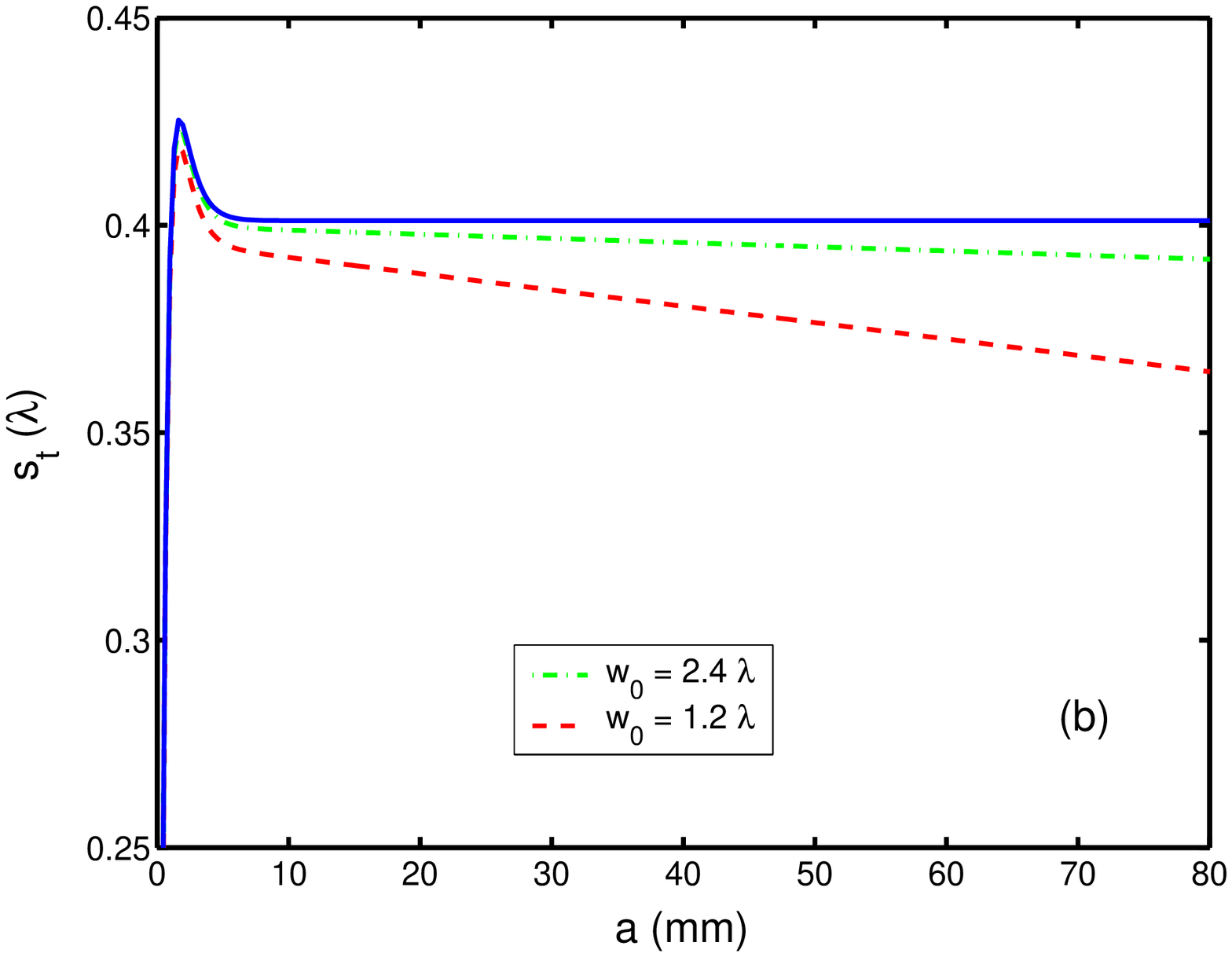}} \caption{(Color
online) Dependence of GH shift (in unit of $\lambda$) on width of
air gap with different beam widths, where (a) $\theta_0=45^{\circ}$
(b) $\theta_0=75^{\circ}$, and other parameters are the same as in
Fig. \ref{fig.2}. The solid corresponds to GH shift in the
first-order approximation, the dashed and dotted curves correspond
to the GH shifts in the second-order approximation.} \label{fig.3}
\end{figure}
\begin{equation}
F_t=F_t^{'}+iF_t^{''}=-i\frac{d}{dk_{y0}}\left(\frac{1}{T}\frac{dT}{dk_y}\right),
\end{equation}
then, with the phase and magnitude of $T(k_y)$. we have
$$
F_t^{'}=\frac{d^2\phi}{d k^2_{y0}},
$$
and
$$
F_t^{''}=-\frac{d^2}{dk^2_{y0}}\ln |T|.
$$
Substituting expression (\ref{transmission series-2}) into Eq.
(\ref{transmitted beam}) using paraxial condition
(\ref{approximation condition}), and neglecting some unimportant
factors, we finally obtain the following field of the transmitted
beam at $x=a$,
\begin{eqnarray}
\label{field of transmitted beam-2} \Psi_t(a,y) & \approx & T_0 \exp
\left[-\frac{1}{2{w_{tf}^2}}\left(y-L'_t+\frac{{\eta_t}
F'_t}{w_{ty}}\right)^2\right] \\\nonumber &\times&
\exp\left[i\left(k_{y0}+\frac{\eta_t}{w_{ty}} \right) y \right],
\end{eqnarray}
where $\eta_t = L''_t/w_{ty}$, $\label{wf-wm} w_{tf} =
\left(w_{ty}^2- i F'_t\right)^{1/2}$, and $w_{ty}^2 = w_y^2+F''_t$
correspond to the angular deflection, focal shift and waist-width
modification, respectively \cite{Li-ZCZ}. The GH shift in this case
can be expressed by,
\begin{equation}
\label{correction} s_t=L_t'-\frac{\eta_tF_t'}{w_{ty}}.
\end{equation}
Obviously, the second term on the right-handed side of Eq.
(\ref{correction}) is a second-order correction, which leads to the
dependence of the GH shift on the beam width. In addition, it also
results in the dependence of the GH shift on the width of air gap in
the opaque barrier limit.

Fig. \ref{fig.3} demonstrates that the GH shift in the second-order
approximation depends on the beam width, where (a)
$\theta_0=45^{\circ}$ (b) $\theta_0=75^{\circ}$, and other
parameters are the same as in Fig. \ref{fig.2}. Compared with Fig.
\ref{fig.2} discussed above, the GH shift becomes dependent on the
width $a$ in the limit of an opaque barrier, due to the second
correction. When the beam width is large, that is, the divergence
angle becomes small, the correction to GH shift can be neglected,
thus for a well-collimated beam the GH shift is in agreement with
that given by the stationary phase method. More importantly, Fig.
\ref{fig.3} shows that the GH shift increases with decreasing the
beam width at $\theta_0=45^{\circ}$, while the GH shift for
$\theta_0=75^{\circ}$ shows a strong decrease with decreasing the
beam width. As shown in Fig. \ref{fig.3} (a), the GH shift becomes
dependent linearly on the width of air gap, because the Fourier
components of the incident beam above the critical angle are
strongly depressed so that the plane wave components just below the
critical angle start to dominate. That is to say, when the incidence
angle is larger than but close to the critical angle, the wave
vector filter is more pronounced for a larger beam width, the
transmission is essentially not tunneling at all, thus the GH shift
increases with increasing the width $a$, as one expects classically.
This also implies the violation of ``Hartman effect" for the quantum
tunneling in time domain \cite{Muga}.

Finally, we have a brief look at the microwave experiment on GH
shifts. It was once argued \cite{Haibel} that the influences of beam
width and incidence angle challenge the current descriptions of the
GH shift in FTIR. Fig. \ref{fig.3} (a) shows the GH shift increases
with deceasing beam dimension corresponding to the beam waist width,
which is agreement with the experimental results in Ref.
\cite{Haibel} where the physical parameters are the same as those in
Fig. \ref{fig.2}. In addition, it is also predicted in Fig.
\ref{fig.3} (b) that the GH shift will decreases with deceasing beam
waist, when the incidence angle is away from the critical angle. In
a word, the improved formula of GH shift (\ref{correction}) with the
modification of beam width can give better understanding of the GH
shift in FTIR theoretically and experimentally.

To summary, we have investigated the GH shifts in FTIR by wave
packet propagation. It is found that the GH shift in the first-order
approximation of the transmission coefficient, which is exactly the
expression of the GH shift obtained by stationary phase method,
approaches the saturation value in two different ways depending on
the angle of incidence. The explicit expression of the GH shift in
the second-order approximation shows the strong dependence on the
beam width. It is further shown that the GH shift with the
second-order correction increases with decreasing the beam width at
the small incident angles, while for the large incident angles the
GH shift reveals a decrease with decreasing the beam width. All
these theoretical results can be applicable to explain the
experiment on GH shifts \cite{Haibel} and offer a hint to the better
understanding of tunneling delay time in FTIR \cite{Winful-Zhang}.

%\section*{Acknowledgments}
This work was supported in part by the National Natural Science
Foundation of China (60806041, 60877055), the Shanghai Rising-Star
Program (08QA14030), the Science and Technology Commission of
Shanghai Municipal (08JC14097), the Shanghai Educational Development
Foundation (2007CG52), and the Shanghai Leading Academic Discipline
Program (S30105). X. Chen is also supported by Programme Juan de la
Cierva of Spanish Ministry of Science and Innovation.


\begin{thebibliography}{99}
\bibitem{Caculation} The neccessay condition can also be derived from $\tau_i>\tau_d$ given in Ref. [28] in the limit of thin barrier.
\end{thebibliography}
\end{document}